\documentclass[aps,prb,twocolumn,letterpaper,superscriptaddress,showpacs]{revtex4-1}
\usepackage{keyval}%
\usepackage{graphicx}
\usepackage{dcolumn}
\usepackage{bm}
\usepackage{color}
\usepackage{amsmath, amssymb}
\usepackage{soul}

\begin{document}

\title{Tight-binding calculations of optical matrix elements for conductivity using non-orthogonal atomic orbitals: Anomalous Hall conductivity in bcc Fe}
\author{Chi-Cheng Lee}
\affiliation{Institute for Solid State Physics, The University of Tokyo, 5-1-5 Kashiwanoha, Kashiwa, Chiba 277-8581, Japan}%
\author{Yung-Ting Lee}
\affiliation{Institute for Solid State Physics, The University of Tokyo, 5-1-5 Kashiwanoha, Kashiwa, Chiba 277-8581, Japan}%
\author{Masahiro Fukuda}
\affiliation{Institute for Solid State Physics, The University of Tokyo, 5-1-5 Kashiwanoha, Kashiwa, Chiba 277-8581, Japan}%
\author{Taisuke Ozaki}
\affiliation{Institute for Solid State Physics, The University of Tokyo, 5-1-5 Kashiwanoha, Kashiwa, Chiba 277-8581, Japan}%
\date{\today}

\begin{abstract}
We present a general formula for the tight-binding representation of momentum matrix elements needed for calculating the conductivity based on 
the Kubo-Greenwood formula using atomic orbitals, which are in general not orthogonal to other orbitals at different sites.  
In particular, the position matrix element is demonstrated to be
important for delivering the exact momentum matrix element. This general formula, applicable to both orthonormal and non-orthonormal bases, solely needs 
the information of the position matrix elements and the ingredients that have already contained in the tight-binding representation.
We then study the anomalous Hall conductivity in the standard example, ferromagnetic bcc Fe, by a first-principles tight-binding Hamiltonian. 
By assuming the commutation relation $\hat{\vec{p}} = (i m_e/\hbar) [\hat{H},\hat{\vec{r}}]$, the obtained frequency-dependent Hall conductivity is 
found to be in good agreement with existing theoretical and experimental results. Better agreement with experiments can be reached by introducing 
a reasonable bandwidth renormalization, evidencing the strong correlation among 3$d$ orbitals in bcc Fe. 
Since a tight-binding Hamiltonian can be straightforwardly 
obtained after finishing a first-principles calculation using atomic basis functions that are generated before the self-consistent calculation, 
the derived formula is particularly useful for those first-principles calculations. 
\end{abstract}

\maketitle

\section{Introduction}

In solid-state physics, the introduction of derivative of Hamiltonian has many advantages. 
One famous example is the Hellmann-Feynman theorem, where the derivative of the energy eigenvalue with respect to a parameter $\lambda$ 
can be calculated via the expectation value of the derivative of Hamiltonian with respect to $\lambda$. In the first-principles calculations 
based on density functional theory\cite{Hohenberg,Kohn}, Hellmann-Feynman theorem 
has been applied for obtaining the forces on nuclei to study ground-state structures
and molecular dynamics\cite{PhysRevLett.55.2471,RevModPhys.64.1045}. Another example is the form of the momentum operator for a periodic system, 
which can be written as the derivative of the Hamiltonian $\hat{H}_u(\vec{k})$ with respect to the crystal momentum $\vec{k}$. Here, $\hat{H}_u(\vec{k})$ is the 
Hamiltonian whose eigenfunctions correspond to the cell-periodic parts of Bloch wave functions. This useful expression has been widely adopted to 
study conductivity, especially in the intrinsic contribution to the anomalous Hall effect that has an intimate relationship to 
the topology of the electronic structures of studied systems\cite{berry1984quantal,PhysRevLett.49.405,PhysRevB.74.195118,RevModPhys.82.1539,RevModPhys.82.1959}.

To study the optical conductivity described by a tight-binding Hamiltonian $\hat{H}$, it is possible to obtain the momentum matrix element 
by just calculating the derivative of the Hamiltonian matrix element with respect to the crystal momentum $\vec{k}$ 
in some specific cases\cite{PhysRevB.47.15500,PhysRevB.51.4940}. Since the properties of basis functions for a tight-binding Hamiltonian obtained
from a fitting procedure or a theoretical model are generally unknown and the derivative has a simple form, for example, 
$t e^{i\vec{k}\cdot\vec{R}}$ to $i\vec{R} t e^{i\vec{k}\cdot\vec{R}}$, having that all the needed ingredients are already self-contained 
in the tight-binding representation is attractive and allows much easier calculations.

However, to study the conductivity with a general set of $\hat{H}$ and bases, one additional term, the position matrix element,
needs to be taken into account\cite{PhysRevB.63.201101}. 
For the case having zero onsite contribution due to orbital symmetry together with the fact that the overlaps 
between intersite orbitals are negligible in the studied system, the position operator sandwiched by atomic orbitals could be neglected, but this approximation
needs to be adopted with caution since the intersite position matrix elements are usually non-negligible. 
The nature of the intra-atomic matrix elements of the position operator can also be understood from 
the non-orthogonality of the atomic orbitals and could play an important role in some cases\cite{PhysRevB.72.125105}. Nevertheless, a general formula for 
the momentum matrix element expressed by the derivative of Hamiltonian matrix element with respect to $\vec{k}$ and the position matrix element  
in the bases of atomic orbitals, which are in general not orthogonal to other orbitals at different sites, has not been derived. 

The frequency-dependent Hall conductivity in bcc Fe has been studied by first-principles calculations within the generalized gradient approximation (GGA)\cite{GGA}, 
where good agreement with experiments is found\cite{PhysRevLett.92.037204,PhysRevB.75.195121}. On the other hand, first-principles studies have also shown that   
bandwidth renormalization needs to be taken into account to compare with the measured quasiparticle bands of bcc Fe in angle-resolved 
photoelectron spectroscopy experiments\cite{PhysRevB.72.155115,PhysRevB.93.205151}. 
It is then interesting to see whether the effect of bandwidth renormalization could give better agreement for the optical conductivity.

In this study, we focus on the tight-binding representation of the momentum matrix elements needed for calculating the optical conductivity 
based on the Kubo-Greenwood formula\cite{kubo1957statistical,greenwood1958boltzmann,PhysRevB.67.235203,allen2006electron,calderin2017kubo}, which 
has been of great interest for not only being able to unveil the excited properties of solids but also the connection to the Berry curvature that
can reveal the topology of the electronic structures\cite{berry1984quantal,PhysRevLett.49.405,PhysRevB.74.195118,RevModPhys.82.1539,RevModPhys.82.1959}. 
In Sec.~\ref{sec:electricalconductivity}, we will derive the general formula for the momentum matrix elements in the bases of atomic orbitals 
without assuming the orthonormal relation among all the orbitals. In Sec.~\ref{sec:bccFe}, the results of frequency-dependent Hall conductivity, 
for a standard example, bcc Fe\cite{PhysRevLett.92.037204,PhysRevB.75.195121}, will be discussed. Finally, a summary is given in Sec.~\ref{sec:summary}.

\section{Optical conductivity}
\label{sec:electricalconductivity}

In this section, we will discuss the optical conductivity using the Kubo-Greenwood formula\cite{allen2006electron} and derive the tight-binding 
representation of the momentum matrix element, $\langle \vec{k}m|\hat{\vec{p}}|\vec{k}n\rangle$, in a non-orthonormal basis. 
While the intraband contribution to the conductivity can solely rely on the knowledge of the tight-binding Hamiltonian 
($\langle 0 M|\hat{H}|\vec{R}N\rangle$) and the overlap matrix ($\langle 0 M|\vec{R}N\rangle$), the interband contribution, 
similar to the case using an orthonormal basis\cite{PhysRevB.63.201101,PhysRevB.74.195118}, 
requires the knowledge of the position matrix element, $\langle 0 M|\hat{\vec{r}}|\vec{R}N\rangle$, to deliver the exact conductivity.

\subsection{Kubo-Greenwood formula}
\label{sec:theory}
The frequency-dependent optical conductivity, $\sigma(\omega)$, expressed by the Kubo-Greenwood formula 
in the bases of Bloch states can be formulated as
\begin{equation}
\sigma_{\alpha\beta}(\omega)=\frac{-i\hbar}{N_k\Omega}\sum_{\vec{k} mn}\Big(\frac{f_{\vec{k}m} - f_{\vec{k}n}}{\epsilon_{\vec{k}m} - \epsilon_{\vec{k}n}}\Big)\frac{\langle \vec{k}m|\hat{j}_{\alpha}|\vec{k}n\rangle\langle \vec{k}n|\hat{j}_\beta|\vec{k}m\rangle}{\omega+\epsilon_{\vec{k}m} - \epsilon_{\vec{k}n}+i\eta}. 
\label{eq:kubo}
\end{equation}
The current operator $\hat{j}_{\alpha}$ can be written as the momentum operator $-\hat{p}_{\alpha}$ in the atomic unit. 
$f_{\vec{k}n}$ denotes the occupation number of the Bloch state $|\vec{k}n\rangle$ 
labeled by the crystal momentum $\vec{k}$ and the band index $n$ (lowercase letters) with the energy $\epsilon_{\vec{k}n}$, where the 
Fermi-Dirac distribution can be applied for $f_{\vec{k}n}$ to introduce the effect of temperature $T$. 
The summation of $\vec{k}$ is over the $k$ points inside the first Brillouin zone and the total number 
of $k$ points is denoted as $N_k$. The parameter $\eta$ means $0^+$ but is a tunable parameter in practice. 
$\Omega$ denotes the volume of the unit cell. In the case of $\epsilon_{\vec{k}m} \rightarrow \epsilon_{\vec{k}n}$
for the degenerate states or the intraband contribution, 
$(f_{\vec{k}m} - f_{\vec{k}n})/(\epsilon_{\vec{k}m} - \epsilon_{\vec{k}n})$ 
should be considered as the derivative of the occupation number with respect to the energy\cite{allen2006electron}, which can be reformulated as
$\partial f(\epsilon_{\vec{k}n})/\partial\epsilon_{\vec{k}n}$ with 
\begin{equation}
f(\epsilon_{\vec{k}n})=\frac{1}{e^{(\epsilon_{\vec{k}n}-\mu)/k_B T}+1},
\end{equation}
where $\mu$ and $k_B$ are the chemical potential and Boltzmann constant, respectively. 

\subsection{Optical matrix element}

To study the optical conductivity using Eq.~\ref{eq:kubo}, the momentum matrix element, 
$\langle \vec{k}m|\hat{\vec{p}}|\vec{k}n\rangle$, needs to be calculated in a basis.
Unlike orthonormal basis functions, the atomic orbital $|\vec{R}N\rangle$ labeled by the lattice vector $\vec{R}$ and 
the orbital index $N$ (capital letters) is in general not orthogonal to the one at a different site. 
So the orthonormal relation for the overlap matrix element
$S_{\vec{R}^\prime M,\vec{R}N}\equiv\langle\vec{R}^\prime M|\vec{R}N\rangle=\delta_{\vec{R}^\prime R}\delta_{MN}$ does not hold in general. 
Consequently, the conductivity described by the tight-binding Hamiltonian represented by such basis functions is expected to require the knowledge
of $S_{\vec{R}^\prime M,\vec{R}N}$. $S_{\vec{R}^\prime M,\vec{R}N}$ is also expected to be important for calculating the momentum matrix elements. 

The energy eigenvalue $\epsilon_{\vec{k}n}$ and the energy eigenstate $|\vec{k}n\rangle$, 
which is expanded by $\sum_{N} C^{\vec{k}n}_N |\vec{k}N\rangle$, can be obtained by solving the generalized eigenvalue problem: 
\begin{equation}
\sum_N H_{MN}(\vec{k}) C^{\vec{k}n}_N = \epsilon_{\vec{k}n} \sum_{N} S_{MN}(\vec{k}) C^{\vec{k}n}_{N},
\label{eq:eq03}
\end{equation}
where $H_{MN}(\vec{k})\equiv\langle\vec{k}M|\hat{H}|\vec{k}N\rangle$ and $S_{MN}(\vec{k})\equiv\langle\vec{k}M|\vec{k}N\rangle$.
The $\langle\vec{k}n|\hat{\vec{p}}|\vec{k}n\rangle$ can be derived by first noting that 
\begin{equation}
\epsilon_{\vec{k}n} = \sum_{MN} C^{\vec{k}n*}_M H_{MN}(\vec{k}) C^{\vec{k}n}_N. 
\label{eq:eq04}
\end{equation}
The expectation value of momentum can be obtained by taking the derivative of the energy with respect to $\vec{k}$.
By utilizing Eq.~\ref{eq:eq03} and  
\begin{align}
\frac{\partial}{\partial \vec{k}} \sum_{MN} C^{\vec{k}n*}_M C^{\vec{k}n}_N S_{MN}(\vec{k}) = 0,
\end{align}
the derivative of Eq.~\ref{eq:eq04} with respect to $\vec{k}$ can be formulated as
\begin{align}
  \frac{\partial \epsilon_{\vec{k}n}}{\partial \vec{k}}
 = \sum_{MN} C^{\vec{k}n *}_M C^{\vec{k}n}_N \Big( \frac{\partial H_{MN}(\vec{k})}{\partial \vec{k}} 
   - \epsilon_{\vec{k}n} \frac{\partial S_{MN}(\vec{k})}{\partial \vec{k}} \Big),  
\label{eq:eq06}
\end{align}
where all the needed information besides the solution of Eq.~\ref{eq:eq03} is the Hamiltonian matrix element, 
$\langle 0 M|\hat{H}|\vec{R}N\rangle$, and the overlap matrix element, $\langle 0 M|\vec{R}N\rangle$. The resulting formula is expected
by considering Hellmann-Feynman theorem for the solution of the generalized eigenvalue problem\cite{PhysRevB.47.12754}. 
Therefore, the intraband contribution, as for the Drude conductivity, can be obtained solely by the knowledge of the tight-binding representation.

We now consider the matrix element for the interband contribution. 
By assuming the commutation relation $\hat{\vec{p}} = (i m_e/\hbar) [\hat{H},\hat{\vec{r}}]$ ($m_e=1$ and $\hbar=1$ in the atomic unit) holds for
the Hamiltonian $\hat{H}$, as derived in Appendix~\ref{sec:derivation}, the momentum operator sandwiched by 
different energy eigenstates can be formulated as
\begin{align}
 & \langle \vec{k}m|\hat{\vec{p}}|\vec{k}n\rangle \nonumber \\
 = & \sum_{MN} C^{\vec{k}m *}_M C^{\vec{k}n}_N \Big( \frac{\partial H_{MN}(\vec{k})}{\partial \vec{k}}-\epsilon_{\vec{k}m} 
   \frac{\partial S_{MN}(\vec{k})}{\partial \vec{k}} \Big) \nonumber \\
 + & i(\epsilon_{\vec{k}m}-\epsilon_{\vec{k}n})\sum_{MN} C^{\vec{k}m *}_M C^{\vec{k}n}_N 
    \sum_{\vec{R}} \langle 0 M|\hat{\vec{r}}|\vec{R}N\rangle e^{i\vec{k}\cdot\vec{R}},
\label{eq:eq07}
\end{align}
where additional information, $\langle 0 M|\hat{\vec{r}}|\vec{R}N\rangle$, is needed to deliver the exact value. 
Although $\langle 0 M|\hat{\vec{r}}|\vec{R}N\rangle$ is origin-dependent, 
the second term of the right-hand side of Eq.~\ref{eq:eq07} is origin-independent as discussed in Appendix~\ref{sec:derivation}.
Importantly, the overlap $\langle 0 M|\vec{R}N\rangle$ diminishes rapidly before $\vec{r}$ goes to infinity.
The importance of such position matrix elements has already been realized for calculating the momentum matrix elements
in an orthonormal basis\cite{PhysRevB.63.201101,PhysRevB.74.195118}. We also note that the diagonal momentum matrix element, Eq.~\ref{eq:eq06}, 
can be reached by Eq.~\ref{eq:eq07}. 

\subsection{Discussions}

First, we note that Eq.~\ref{eq:eq07}, as shown in Appendix~\ref{sec:derivation}, 
can be alternatively expressed as
\begin{align}
 & \langle \vec{k}m|\hat{\vec{p}}|\vec{k}n\rangle \nonumber \\
 = & i\sum_{MN} C^{\vec{k}m *}_M C^{\vec{k}n}_N \sum_{\vec{R}} \langle0 M|\hat{H}|\vec{R}N\rangle \vec{R} e^{i\vec{k}\cdot\vec{R}} \nonumber \\
 + & i\epsilon_{\vec{k}m}\sum_{MN} C^{\vec{k}m *}_M C^{\vec{k}n}_N 
   \Big(\sum_{\vec{R}} \langle 0 N|\hat{\vec{r}}|\vec{R}M\rangle e^{i\vec{k}\cdot\vec{R}}\Big)^* \nonumber \\
 - & i\epsilon_{\vec{k}n}\sum_{MN} C^{\vec{k}m *}_M C^{\vec{k}n}_N \sum_{\vec{R}} 
   \langle 0 M|\hat{\vec{r}}|\vec{R}N\rangle e^{i\vec{k}\cdot\vec{R}},
\label{eq:eq011}
\end{align}
where besides the solution of the generalized eigenvalue problem, only the Hamiltonian matrix elements and position matrix elements 
are required for delivering the exact momentum matrix elements. The overlap matrix element, which was expected to play an important
role in a non-orthonormal basis, is not needed explicitly. 

Since Eq.~\ref{eq:eq011} is also applicable to an orthonormal basis, the information of position matrix elements
is needed to deliver the exact momentum matrix elements even in an orthonormal basis, which can also be found in Eq.~\ref{eq:eq07}:
\begin{align}
 & \langle \vec{k}m|\hat{\vec{p}}|\vec{k}n\rangle \nonumber \\
 = & \sum_{MN} C^{\vec{k}m *}_M C^{\vec{k}n}_N \frac{\partial H_{MN}(\vec{k})}{\partial \vec{k}} \nonumber \\
 + & i(\epsilon_{\vec{k}m}-\epsilon_{\vec{k}n})\sum_{MN} C^{\vec{k}m *}_M C^{\vec{k}n}_N 
    \sum_{\vec{R}} \langle 0 M|\hat{\vec{r}}|\vec{R}N\rangle e^{i\vec{k}\cdot\vec{R}}  
\end{align}
for an orthonormal basis.
To study $\sigma_{\alpha\beta}$ in the static limit for analyzing the quantized Hall conductance, the momentum matrix element is commonly discussed via 
the cell-periodic wave function $|u_{\vec{k}n}\rangle$. Recall that the energy eigenstate $|kn\rangle$ can be written in the Bloch form:
\begin{align}
\Psi_{\vec{k}n}(\vec{r})= e^{i\vec{k}\cdot\vec{r}} u_{\vec{k}n}(\vec{r}),
\end{align}
where the cell-periodic wave function $|u_{\vec{k}n}\rangle$ satisfies
\begin{align}
\hat{H}_u(\vec{k}) |u_{\vec{k}n}\rangle = \epsilon_{\vec{k}n} |u_{\vec{k}n}\rangle.
\label{eq:eq012}
\end{align}
Thanks to $|u_{\vec{k}n}\rangle$, the momentum matrix element is
\begin{align}
\langle \vec{k}m|\hat{\vec{p}}|\vec{k}n\rangle = \langle u_{\vec{k}m}| \frac{\partial \hat{H}_u (\vec{k})}{\partial \vec{k}} |u_{\vec{k}n}\rangle.
\label{eq:eq013}
\end{align}
The famous TKNN (Thouless, Kohmoto, Nightingale, and Nijs) formula\cite{PhysRevLett.49.405} for studying $\sigma_{\alpha\beta}(\omega\rightarrow 0)$ can be derived by utilizing Eq.~\ref{eq:eq013}, and then
be used to connect to the Berry curvature\cite{berry1984quantal,PhysRevLett.49.405,PhysRevB.74.195118,RevModPhys.82.1539,RevModPhys.82.1959}. 
Since the momentum matrix elements should agree with each other calculated from both methods associated with two different Hamiltonians,
the $k$-dependent position matrix element via Fourier transform as a correction term to $\partial H_{MN}(\vec{k})/\partial \vec{k}$ is essential for delivering the same
result of TKNN formula using the Kubo-Greenwood formula by summing all of the eigenstates of $\hat{H}$. 
It is worth mentioning that TKNN formula requires solely the knowledge of occupied bands, 
which can also be well described by first-principles calculations using atomic basis functions.

Another issue is the commutation relation $\hat{\vec{p}}=i[\hat{H},\hat{\vec{r}}]$, which is assumed to be valid in deriving Eq.~\ref{eq:eq07}.
For the case where such a relation does not hold, the momentum should be obtained by $\hat{\vec{p}}=i[\hat{H}-\hat{H}^\prime,\hat{\vec{r}}]$, where
the commutator $[\hat{H}^\prime,\hat{\vec{r}}]$ must be taken into account as a correction term to $[\hat{H},\hat{\vec{r}}]$ for delivering 
the exact value of the momentum matrix element. An example of $\hat{H}^\prime$ is the spin-orbit coupling term discussed elsewhere\cite{PhysRevB.47.15500} 
although the correction is estimated to be small. In first-principles calculations, full potentials are commonly replaced by pseudopotentials, and
the non-local form, $\sum_{lm} |lm\rangle V_{lm} \langle lm|$, which does not commute with $\hat{\vec{r}}$ in general, is also commonly adopted.
The error due to the use of pseudopotentials in calculating momentum matrix elements could be large and depends on the studied 
systems\cite{PhysRevB.44.13071,PhysRevB.56.14985}. To reach the solution of a full-potential calculation, for example, 
$-i\langle \vec{k}n^{full} |\partial / \partial \hat{\vec{r}} | \vec{k}n^{full} \rangle$, from the pseudopotential solution, 
$-i\langle \vec{k}n^{pseudo} |\partial / \partial \hat{\vec{r}} | \vec{k}n^{pseudo} \rangle$, the addition of projector augmented wave can 
recover the difference in the wave functions\cite{calderin2017kubo}. It can be found that the correction to either the commutation relation or 
the wave functions could require knowledge way beyond a simple tight-binding representation. Therefore, we propose that Eq.~\ref{eq:eq07} can serve as 
a good starting point to study the optical conductivity. As we will show in Sec.~\ref{sec:bccFe}, the calculated frequency-dependent conductivity 
in bcc Fe using Eq.~\ref{eq:eq07} is in good agreement with the reported theoretical results\cite{PhysRevLett.92.037204,PhysRevB.75.195121}.

Finally, it should be noted that a limited finite number of atomic basis functions is insufficient to describe a first-principles Hamiltonian.
While the Hamiltonian represented by the atomic orbitals could usually give a good description of occupied bands, it is difficult to reproduce accurate 
unoccupied bands up to a high energy. Therefore, incomplete atomic orbitals could lead to inaccurate frequency-dependent conductivity. 
For the study of conductivity in bcc Fe, 13 atomic orbitals locating at each atomic site are found to be enough to describe the frequency range 
we will study in Sec.~\ref{sec:bccFe}.
The position operator sandwiched by the energy eigenstates expressed by the atomic orbitals is also expected to deviate from the accurate position matrix element 
due to the incomplete basis set even without adopting pseudopotentials. However, Eq.~\ref{eq:eq07} is still useful for describing the momentum matrix elements
as long as the Hamiltonian represented by the finite number of atomic orbitals can well describe the studied physical quantities. For example, the diagonal 
momentum matrix element shown in Eq.~\ref{eq:eq07} can deliver accurate Fermi velocity, which is associated with 
$\partial \epsilon_{\vec{k}n}/\partial \vec{k}$ at the Fermi energy as confirmed by the relationship between
Eq.~\ref{eq:eq07} and Eq.~\ref{eq:eq06}. Eq.~\ref{eq:eq07} and Eq.~\ref{eq:eq06} are obtained from $\hat{\vec{p}}=i[\hat{H},\hat{\vec{r}}]$ and
the generalized eigenvalue problem (Eq.~\ref{eq:eq03}), respectively. Obviously, the same $\partial \epsilon_{\vec{k}n}/\partial \vec{k}$ can be 
reached by the calculations using different kinds of approaches and is measurable by angle-resolved photoelectron experiments.
In contrast, a direct calculation of $-i\langle \vec{k}n|\partial/\partial {\vec{r}}| \vec{k}n\rangle$ could deviate from 
$\partial \epsilon_{\vec{k}n}/\partial \vec{k}$ noticeably depending on the adopted bases and pseudopotentials.

\section{Anomalous Hall conductivity in bcc Fe}
\label{sec:bccFe}

For a benchmark calculation, we focus on the anomalous Hall conductivity for the standard example, bcc Fe. The intrinsic contribution to 
the anomalous Hall conductivity in ferromagnetic bcc Fe has been studied by first-principles calculations using 
Kubo-Greenwood formula\cite{PhysRevLett.92.037204}, where the results are in good agreement with other theoretical calculations and 
experiments\cite{PhysRev.156.637,krinchik1968magneto,PhysRevB.51.12633,PhysRevB.53.3692}. In this section, we will show the resulting  
frequency-dependent Hall conductivity using Eq.~\ref{eq:eq07} and compare with the reported conductivity\cite{PhysRevLett.92.037204,PhysRevB.75.195121}. 
How to reach better agreement between theory and experiment will also be discussed.

\subsection{Computational details}

The first-principles calculations were performed using the OpenMX code,\cite{openmx} 
where the GGA, the norm-conserving relativistic non-local pseudopotentials, and optimized pseudo-atomic 
basis functions were adopted\cite{GGA,Theurich,Morrison,Ozaki}. The spin-orbit coupling was incorporated through $j$-dependent pseudopotentials\cite{Theurich}.
Two, two, and one optimized radial functions were allocated for the $s$, $p$, and $d$ orbitals, respectively, for the Fe atom with a
cutoff radius of 6 Bohr. A cutoff energy of 340 Ha was used for numerical integrations and for the solution of the Poisson equation.
The $30\times30\times30$ $k$-point sampling was adopted for the experimental lattice constant, 2.87 \AA. 
After the self-consistent calculation was done, a tight-binding Hamiltonian in the bases of the 13 adopted 
pseudo-atomic orbitals per Fe atom was obtained and used in calculating the frequency-dependent conductivity using Eqs.~\ref{eq:kubo} and \ref{eq:eq07},
where $\eta =0.05eV$ and a $150\times150\times150$ $k$-mesh were chosen. The electronic temperature was set to 300 K for both of
the first-principles and conductivity calculations. The magnetization direction is along the $z$ direction.

\subsection{Results}

\begin{figure}[tbp]
\includegraphics[width=1.00\columnwidth,clip=true,angle=0]{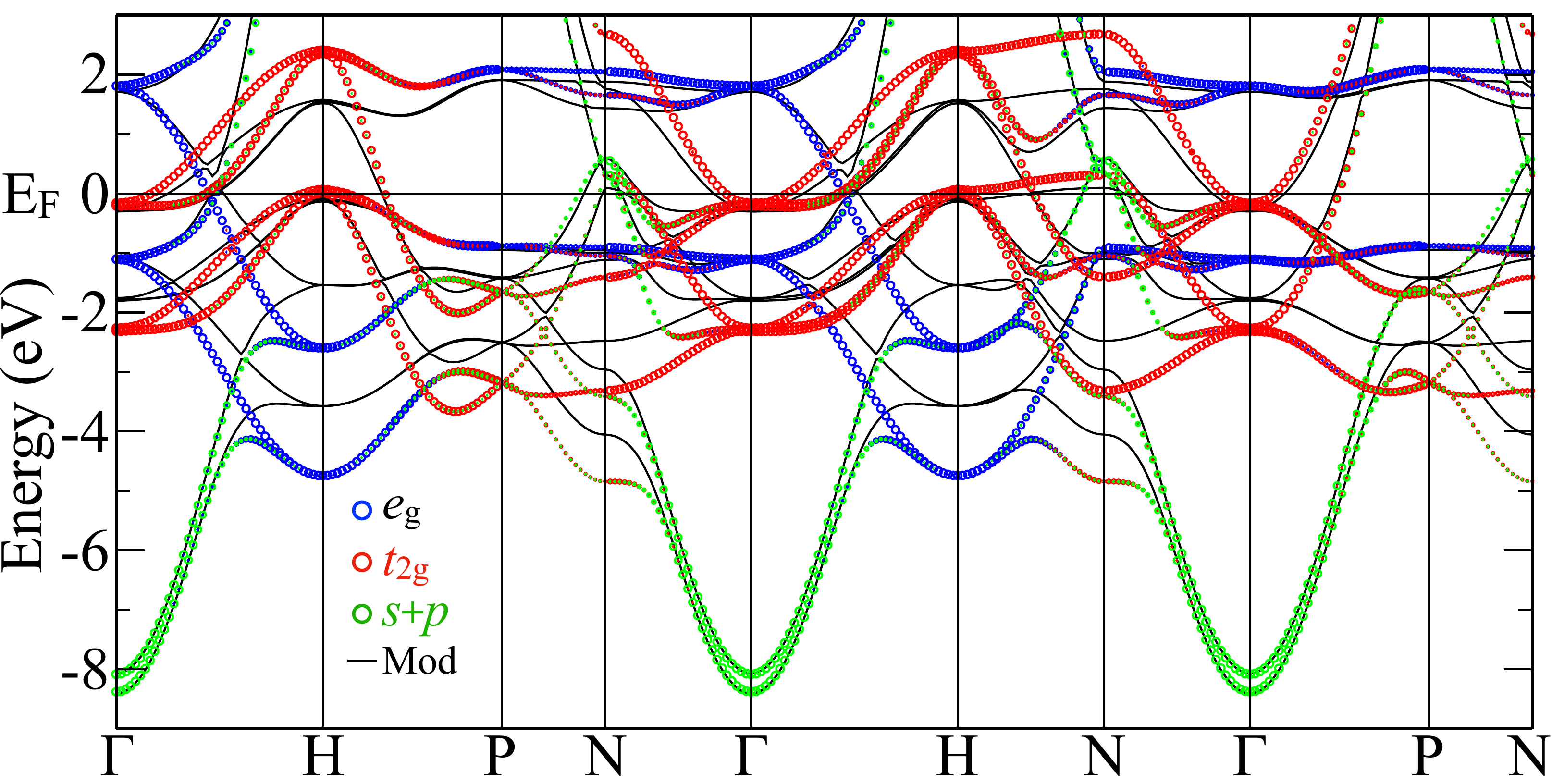}
\caption{First-principles band structure of ferromagnetic bcc Fe obtained by adopting 13 pseudo-atomic orbitals per Fe atom with spin-orbit coupling. 
Only the bands near the Fermi energy, which is shifted to zero as denoted by $E_F$, are shown. Blue and red circles indicate the contribution of 
Fe 3$d$ $e_g$ and $t_{2g}$ orbitals, respectively. The other contribution ($s+p$) is presented by green circles. The modified band structure (Mod) 
with rescaled hopping integrals between $d$ orbitals (80$\%$) and adjusted onsite energies is presented by the black curves.
}
\label{fig:band}
\end{figure}

The band structure of ferromagnetic bcc Fe near the Fermi energy including spin-orbit coupling is shown in Fig.~\ref{fig:band}, where 
the $e_g$, $t_{2g}$, and $s+p$ orbital contributions are presented by blue, red, and green circles, respectively. 
Two sets of $e_g$ or $t_{2g}$ bands with similar dispersion separated by a large gap ($>$ 2 eV at $\Gamma$) can be clearly observed 
and recognized as the spin-up and spin-down bands before being coupled by the spin-orbit coupling. 
The band structure is consistent with the reported one\cite{PhysRevLett.92.037204}, and therefore the same conductivity is expected to be obtained within
density functional theory. Since our formula is based on the Kubo-Greenwood formula involving a 
summation over all the 13 non-orthogonal pseudo-atomic orbitals per Fe atom, a less efficient computation 
compared with the method using Wannier functions is expected\cite{PhysRevB.75.195121}. But it should be noted that the pseudo-atomic orbitals are 
generated before the computation of electronic structure of bcc Fe in comparison with Wannier functions, which need to be constructed after
finishing first-principles calculations. All the needed ingredients for Eq.~\ref{eq:eq07} are straightforwardly obtained 
in our first-principles calculations using the pseudo-atomic basis functions, and our tight-binding Hamiltonian shares the same advantage of 
efficiently calculating the eigenstates at a dense grid of $k$ points needed for describing the Fermi surface of bcc Fe in comparison with the first-principles plane-wave calculations. 

\begin{figure}[tbp]
\includegraphics[width=0.95\columnwidth,clip=true,angle=0]{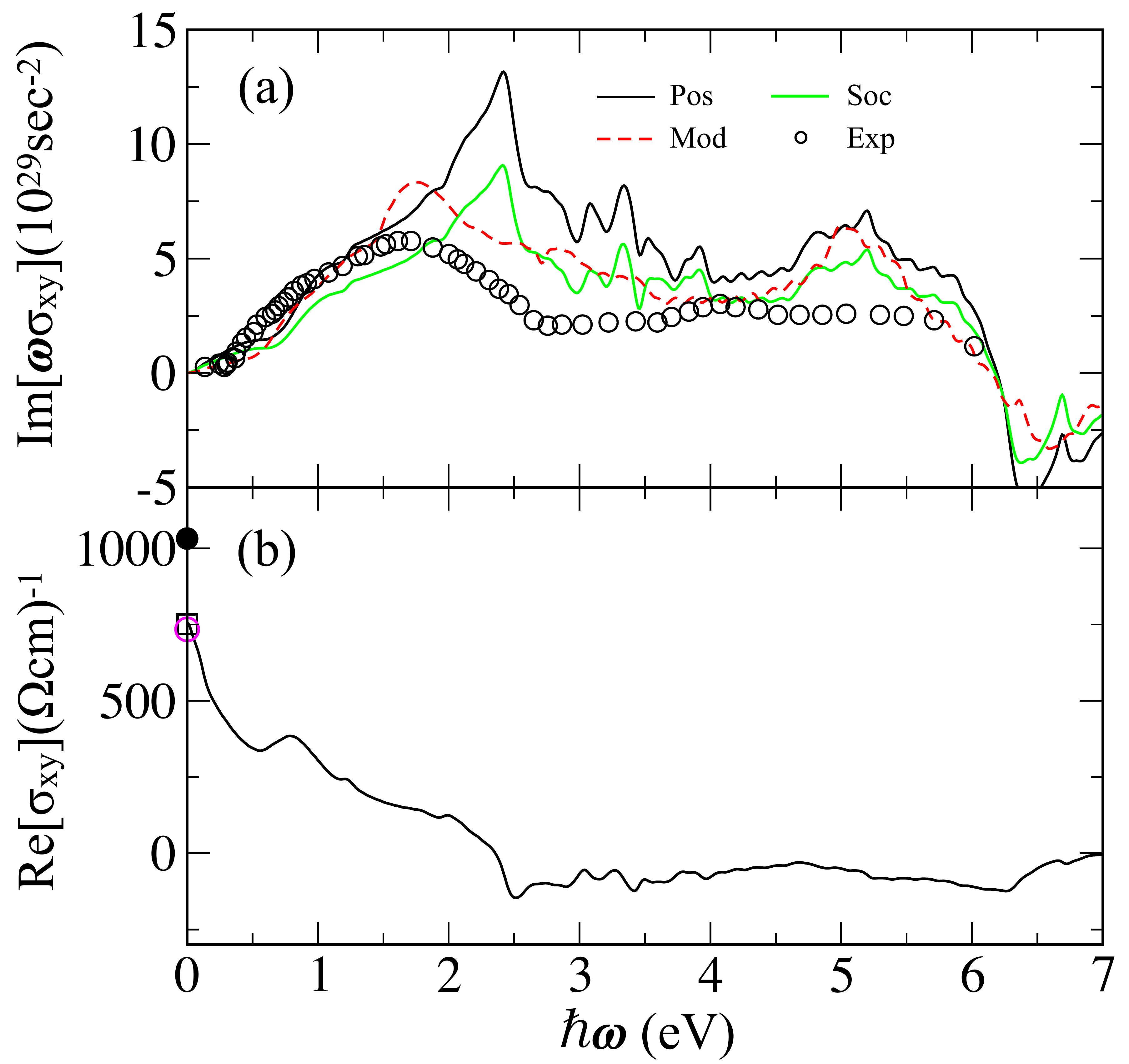}
\caption{Frequency-dependent Hall conductivity in bcc Fe.
(a) The imaginary part of $\omega \sigma_{xy}$, the magnetic circular dichroism spectrum,
is compared with the experimental data presented by open circles from Ref.\citenum{krinchik1968magneto} (Exp) as reproduced in Ref.\citenum{PhysRevLett.92.037204} 
and Ref.\citenum{PhysRevB.75.195121}. The black curve shows the result calculated using 
Eqs.~\ref{eq:kubo} and \ref{eq:eq07} (Pos), and the green curve shows the result by allowing only 70$\%$ strength of the
spin-orbit coupling in the $d$ orbitals (Soc). The result obtained from the modified band structure with rescaled hopping integrals between $d$ orbitals (80$\%$) and 
adjusted onsite energies is presented by the red dashed curve (Mod).
(b) The real part of $\sigma_{xy}$ is shown and compared with the 
dc experiment value\cite{PhysRev.156.637} (solid circle) and the theoretical results for 0 K (open square) and 300 K (open circle), 
as discussed in Ref.\citenum{PhysRevLett.92.037204}. 
}
\label{fig:Hall}
\end{figure}

The calculated frequency-dependent Hall conductivity using Eqs.~\ref{eq:kubo} and \ref{eq:eq07} is presented by the black curves in Fig.~\ref{fig:Hall}.
First, we compare the spectrum of magnetic circular dichroism, which corresponds to the imaginary part of 
$\omega \sigma_{xy}$, with the available theoretical and experimental results. As shown in Fig.~\ref{fig:Hall} (a), 
the curve delivered by Eq.~\ref{eq:eq07} is consistent with other first-principles calculations\cite{PhysRevLett.92.037204,PhysRevB.75.195121}, 
where the spectrum is in good agreement with experiment\cite{krinchik1968magneto} up to about 1.7 eV.
The details of the first-principles results at higher frequencies\cite{PhysRevLett.92.037204,PhysRevB.75.195121}, such as the
prominent peak at $\sim$2.4 eV and a big drop around 6.5 eV, are also well reproduced in our calculations. 
The real part of the Hall conductivity, as shown in Fig.~\ref{fig:Hall} (b),
also agrees very well with the reported one\cite{PhysRevLett.92.037204}, such as the dc limit result and the dip at $\sim$2.5 eV. 
Overall, our calculated real and imaginary parts of Hall conductivity, which should satisfy a Kramers-Kronig relation, 
using Eq.~\ref{eq:eq07} are in good agreement with the reported theoretical and 
experimental data\cite{PhysRevLett.92.037204,PhysRevB.75.195121,PhysRev.156.637,krinchik1968magneto,PhysRevB.51.12633,PhysRevB.53.3692}. 
We attribute the small errors to the hard Fe pseudopotential that is close to the full potential in our calculations. 

Although the general behavior of the conductivity can be described by GGA, the calculated intensity shown in
Fig.~\ref{fig:Hall} (a) is overall higher than the experimental data at energy $> 1.7$ eV, which is consistent with the previous 
findings\cite{PhysRevLett.92.037204,PhysRevB.75.195121}. Since the intensity is related to the strength
of spin-orbit coupling, we have calculated the magnetic circular dichroism spectrum by allowing only 70$\%$ strength 
of the spin-orbit coupling for the $d$ orbitals in the self-consistent calculation. 
As expected, the suppressed spin-orbit coupling 
lowers the intensity towards the experimental data as shown by the green curve in Fig.~\ref{fig:Hall} (a). However, the calculated prominent peak 
still remains at $\sim$2.4 eV, which deviates from the experimental data having the highest intensity at $\sim$1.7 eV, 
and it seems unlikely to reach the 30$\%$ error in the spin-orbit coupling.
This suggests that a more complete description of many-body Coulomb interactions is needed even in the metallic bcc Fe.
The peak at 2.4 eV can be understood from the large number of unoccupied $e_g$ states around 2 eV as shown in Fig.~\ref{fig:band}
that provides a channel for electrons to excite to. The energy of the flatter occupied $e_g$ states corresponding to larger density of states 
can be found to be lower than -0.4 eV, and 
therefore the occupied $t_{2g}$ states have also largely contributed to the peak at 2.4 eV.
To simultaneously lower the overall intensity and shift the highest intensity in the calculated spectrum,
the Coulomb correlations beyond GGA among all of the Fe $3d$ orbitals should be taken into account. 

To compare with the measured bands of bcc Fe in the angle-resolved photoelectron spectroscopy experiments, significant bandwidth renormalization
needs to be introduced to the first-principles band structures\cite{PhysRevB.72.155115,PhysRevB.93.205151}. The bandwidth renormalization
is a signature of strong Coulomb correlations and can be described by the Gutzwiller density functional theory, 
which has revealed a large bandwidth reduction in bcc Fe, for example, by $36\%$ at the $H$ point\cite{PhysRevB.72.155115,PhysRevB.93.205151}. 
We therefore have rescaled the hopping integrals between the $d$ orbitals by 80$\%$ and adjusted the onsite energies of $e_g$ and $t_{2g}$ orbitals to 
reach consistent $\Gamma$-point band energies calculated by the Gutzwiller density functional theory\cite{PhysRevB.93.205151}. 
The modified band structure is presented by the black curves in Fig.~\ref{fig:band}. We note that this simple modification of our tight-binding Hamiltonian 
cannot reproduce the same Fermi surface as the ones calculated by GGA and the Gutzwiller density functional theory, and the modification gives a slightly 
electron-doped system ($\sim0.07 e$) using the GGA chemical potential. But the modified tight-binding Hamiltonian can reflect the effect of 
bandwidth renormalization in the calculated spectrum of magnetic circular dichroism. The result is presented by the red dashed curve in Fig.~\ref{fig:Hall} (a), 
where the lowered overall intensity and the shifted highest intensity towards the experimental data can be identified. This suggests that 
the magnetic circular dichroism experiment has also evidenced the effect of strong Coulomb correlations in bcc Fe.
We expect that an even better improvement can be achieved by considering the many-body Coulomb interactions in the electron-hole channel, which 
is beyond the scope of this study and should be left for a future work.

\section{Summary}
\label{sec:summary}

The tight-binding representation of momentum matrix elements for calculating the optical conductivity based on the Kubo-Greenwood formula using 
the bases of atomic orbitals, where the orthonormal relation is not assumed, is derived. 
To reach the exact value of the momentum matrix element in the tight-binding representation,
the $k$-dependent position matrix element via Fourier transform needs to be taken into account, which is also needed in an orthonormal basis as well. 
For the tight-binding parameters obtained from a fitting procedure, the position matrix elements are unknown due to lacking the knowledge of the basis functions and 
must be parameterized. For the case the tight-binding Hamiltonian is obtained from first-principles calculations using atomic basis functions, 
the position matrix elements can be easily calculated since the atomic basis functions are generated at the step of generating the pseudopotentials.
Once the geometrical structure is determined, the computational effort for calculating the position matrix elements is similar to the calculation 
of overlap matrix elements. 

Although the number of pseudo-atomic orbitals in first-principles calculations is commonly larger than that of the energy-resolved Wannier functions, 
they share the same advantage of calculating the eigenstates at a dense grid of $k$ points by diagonalizing a tight-binding Hamiltonian in comparison with 
the plane-wave calculations. Upon finishing self-consistent first-principles calculations using atomic basis functions, 
such as those implemented in SIESTA\cite{0953-8984-14-11-302}, Conquest\cite{GILLAN200714}, FHI-AIMS\cite{BLUM20092175}, CP2K\cite{hutter2014cp2k}, 
and Atomistix ToolKit\cite{quantumwise}, 
a tight-binding Hamiltonian is straightforwardly obtained as well as the other needed ingredients for Eq.~\ref{eq:eq07}, and therefore can benefit from our formula. 
We have studied the frequency-dependent Hall conductivity in ferromagnetic bcc Fe and demonstrated that the results are in good agreement with 
the reported theoretical and experimental data. By introducing a reasonable bandwidth renormalization by simply rescaling the hopping integrals, 
better agreement with experiments can be reached, evidencing the effect of strong correlation among 3$d$ orbitals in bcc Fe. 
We therefore propose that the derived formula, which is applicable to a non-orthogonal basis, 
is useful for studying the optical conductivity using the tight-binding representation.

\begin{acknowledgments}
This work was supported by Priority Issue (creation of new functional devices and high-performance materials to support next-generation industries) to be tackled by using Post `K' Computer, Ministry of Education, Culture, Sports, Science and Technology, Japan. 
\end{acknowledgments}

\appendix

\section{Derivation of momentum matrix element in non-orthogonal basis}
\label{sec:derivation}
The energy eigenstate $|\vec{k}n\rangle$ can be expanded by the coefficient of linear combination of atomic orbital, $C^{\vec{k}n}_N$, 
and $|\vec{k}N\rangle$ can be expressed by $|\vec{R}N\rangle$ via Fourier transform: 
\begin{align}
|\vec{k}n\rangle = & \sum_{N} C^{\vec{k}n}_N |\vec{k}N\rangle \nonumber \\
                 = & \sum_{\vec{R}N} \frac{e^{i\vec{k}\cdot \vec{R}}}{\sqrt{N_k}} C^{\vec{k}n}_N |\vec{R}N\rangle.
\label{eq:eq1}
\end{align}
The momentum matrix element $\langle \vec{k}m|\hat{\vec{p}}|\vec{k}n\rangle$ becomes:
\begin{align}
\langle \vec{k}m|\hat{\vec{p}}|\vec{k}n\rangle = & i \langle \vec{k}m|(\hat{H}\hat{\vec{r}} - \hat{\vec{r}} \hat{H})|\vec{k}n\rangle \nonumber \\
  = & \frac{i}{N_k}\sum_{\vec{R}N\vec{R}^\prime M} C^{\vec{k}m *}_M C^{\vec{k}n}_N e^{i\vec{k}\cdot(\vec{R}-\vec{R}^\prime)} \nonumber \\
    & \times \langle\vec{R}^\prime M|(\hat{H}\hat{\vec{r}} - \hat{\vec{r}} \hat{H})|\vec{R}N\rangle,
\end{align}
where $m_e=1$ and $\hbar=1$ have been adopted for the commutation relation $\hat{\vec{p}} = i m_e/\hbar [\hat{H},\hat{\vec{r}}]$. 
By inserting the identity operator,
\begin{align}
\hat{I}=\sum_{\vec{R}^{\prime\prime}M^\prime \vec{R}^{\prime\prime\prime}N^{\prime}} |\vec{R}^{\prime\prime}M^\prime\rangle 
        S^{-1}_{\vec{R}^{\prime\prime}M^\prime,\vec{R}^{\prime\prime\prime}N^{\prime}}\langle \vec{R}^{\prime\prime\prime}N^{\prime}|,
\end{align}
we obtain
\begin{align}
&\langle \vec{k}m|\hat{\vec{p}}|\vec{k}n\rangle \nonumber \\
 = & \frac{i}{N_k}\sum_{\vec{R}N\vec{R}^\prime M\vec{R}^{\prime\prime}M^\prime \vec{R}^{\prime\prime\prime}N^{\prime}} C^{\vec{k}m *}_M C^{\vec{k}n}_N e^{i\vec{k}\cdot(\vec{R}-\vec{R}^\prime)} \nonumber \\
 & \times \langle\vec{R}^\prime M|\hat{H}|\vec{R}^{\prime\prime}M^\prime\rangle 
        S^{-1}_{\vec{R}^{\prime\prime}M^\prime,\vec{R}^{\prime\prime\prime}N^{\prime}}\langle \vec{R}^{\prime\prime\prime}N^{\prime}|
 \hat{\vec{r}} |\vec{R}N\rangle \nonumber \\
-  & \frac{i}{N_k}\sum_{\vec{R}N\vec{R}^\prime M\vec{R}^{\prime\prime}M^\prime \vec{R}^{\prime\prime\prime}N^{\prime}} C^{\vec{k}m *}_M C^{\vec{k}n}_N e^{i\vec{k}\cdot(\vec{R}-\vec{R}^\prime)} \nonumber \\
 & \times \langle\vec{R}^\prime M|\hat{\vec{r}}|\vec{R}^{\prime\prime}M^\prime\rangle 
        S^{-1}_{\vec{R}^{\prime\prime}M^\prime,\vec{R}^{\prime\prime\prime}N^{\prime}}\langle \vec{R}^{\prime\prime\prime}N^{\prime}|
 \hat{H}|\vec{R}N\rangle.
\label{eq:eq5}
\end{align}
Considering the translational symmetry, one can get
\begin{align}
\langle \vec{R}^{\prime\prime\prime}N^{\prime}|\hat{\vec{r}} |\vec{R}N\rangle = &
   \langle \vec{R}^{\prime\prime\prime} -\vec{R},N^\prime|(\hat{\vec{r}}+\vec{R})|0 N\rangle \nonumber \\
 = & \langle \vec{R}^{\prime\prime\prime} -\vec{R},N^\prime|\hat{\vec{r}}|0 N\rangle \nonumber \\     
 + & \vec{R} S_{\vec{R}^{\prime\prime\prime}N^\prime,\vec{R}N} 
\end{align}
and
\begin{align}
\langle \vec{R}^{\prime}M|\hat{\vec{r}} |\vec{R}^{\prime\prime}M^\prime\rangle = &
   \langle 0 M|(\hat{\vec{r}}+\vec{R}^\prime)|\vec{R}^{\prime\prime}-\vec{R}^\prime,M^\prime\rangle \nonumber \\
 = & \langle 0 M|\hat{\vec{r}}|\vec{R}^{\prime\prime}-\vec{R}^\prime,M^\prime\rangle \nonumber \\     
 + & \vec{R}^\prime S_{\vec{R}^\prime M,\vec{R}^{\prime\prime} M^\prime}. 
\end{align}
By further noting that 
\begin{align}
\sum_j S_{ij} S^{-1}_{jk} = \sum_j S^{-1}_{ij} S_{jk} = \delta_{ik},
\label{eq:eq7}
\end{align}
Eq.~\ref{eq:eq5} can be rewritten as
\begin{align}
&\langle \vec{k}m|\hat{\vec{p}}|\vec{k}n\rangle \nonumber \\
 = & \frac{i}{N_k}\sum_{\vec{R}N\vec{R}^\prime M} C^{\vec{k}m *}_M C^{\vec{k}n}_N e^{i\vec{k}\cdot(\vec{R}-\vec{R}^\prime)}(\vec{R}-\vec{R}^\prime)
    \langle\vec{R}^\prime M|\hat{H}|\vec{R}N\rangle \nonumber \\
 + & \frac{i}{N_k}\sum_{\vec{R}N\vec{R}^\prime M\vec{R}^{\prime\prime}M^\prime \vec{R}^{\prime\prime\prime}N^{\prime}} C^{\vec{k}m *}_M C^{\vec{k}n}_N e^{i\vec{k}\cdot(\vec{R}-\vec{R}^\prime)} \nonumber \\
 & \times \langle\vec{R}^\prime M|\hat{H}|\vec{R}^{\prime\prime}M^\prime\rangle 
        S^{-1}_{\vec{R}^{\prime\prime}M^\prime,\vec{R}^{\prime\prime\prime}N^{\prime}}\langle \vec{R}^{\prime\prime\prime}-\vec{R},N^{\prime}|
 \hat{\vec{r}} |0 N\rangle \nonumber \\
-  & \frac{i}{N_k}\sum_{\vec{R}N\vec{R}^\prime M\vec{R}^{\prime\prime}M^\prime \vec{R}^{\prime\prime\prime}N^{\prime}} C^{\vec{k}m *}_M C^{\vec{k}n}_N e^{i\vec{k}\cdot(\vec{R}-\vec{R}^\prime)} \nonumber \\
 & \times \langle 0 M|\hat{\vec{r}}|\vec{R}^{\prime\prime}-\vec{R}^\prime,M^\prime\rangle 
        S^{-1}_{\vec{R}^{\prime\prime}M^\prime,\vec{R}^{\prime\prime\prime}N^{\prime}}\langle \vec{R}^{\prime\prime\prime}N^{\prime}|
 \hat{H}|\vec{R}N\rangle.
\label{eq:eq8}
\end{align}

The first term of the right-hand side of Eq.~\ref{eq:eq8} can be recognized as the derivative of the Hamiltonian matrix element with respect to $\vec{k}$:
\begin{align}
\frac{\partial}{\partial \vec{k}} (\langle\vec{k}M|\hat{H}|\vec{k}N\rangle) = & 
\sum_{\vec{R}\vec{R}^\prime} \frac{\partial}{\partial \vec{k}} \frac{e^{i\vec{k}\cdot(\vec{R}-\vec{R}^\prime)}}{N_k} \langle\vec{R}^\prime M|\hat{H}|\vec{R}N\rangle \nonumber \\
 = & \sum_{\vec{R}\vec{R}^\prime} i(\vec{R}-\vec{R}^\prime)\frac{e^{i\vec{k}\cdot(\vec{R}-\vec{R}^\prime)}}{N_k} \langle\vec{R}^\prime M|\hat{H}|\vec{R}N\rangle.
\end{align}
The second and third terms on the right-hand side of Eq.~\ref{eq:eq8} can be simplified by considering Eq.~\ref{eq:eq1}, Eq.~\ref{eq:eq7}, and
\begin{align}
\langle\vec{k}m|\hat{H}=\langle\vec{k}m| \epsilon_{\vec{k}m} 
\label{eq:eq10}
\end{align}
or
\begin{align}
\hat{H}|\vec{k}n\rangle=\epsilon_{\vec{k}n} |\vec{k}n\rangle.
\label{eq:eq11}
\end{align}
The momentum matrix element can then be reformulated as 
\begin{align}
 & \langle \vec{k}m|\hat{\vec{p}}|\vec{k}n\rangle \nonumber \\
 = & \sum_{MN} C^{\vec{k}m *}_M C^{\vec{k}n}_N \frac{\partial}{\partial \vec{k}} (\langle\vec{k}M|\hat{H}|\vec{k}N\rangle) \nonumber \\
 + & \frac{i\epsilon_{\vec{k}m}}{N_k}\sum_{\vec{R}N\vec{R}^\prime M} C^{\vec{k}m *}_M C^{\vec{k}n}_N e^{i\vec{k}\cdot(\vec{R}-\vec{R}^\prime)} 
     \langle \vec{R}^\prime -\vec{R},M|\hat{\vec{r}}|0 N\rangle  \nonumber \\
 - & \frac{i\epsilon_{\vec{k}n}}{N_k}\sum_{\vec{R}N\vec{R}^\prime M} C^{\vec{k}m *}_M C^{\vec{k}n}_N e^{i\vec{k}\cdot(\vec{R}-\vec{R}^\prime)}
     \langle 0 M|\hat{\vec{r}}|\vec{R}-\vec{R}^\prime,N\rangle.
\end{align}
After considering translational symmetry, we obtain an expression by assuming that $\langle0 M|\hat{H}|\vec{R}N\rangle$ and
$\langle 0 M|\hat{\vec{r}}|\vec{R}N\rangle$ are known:
\begin{align}
 & \langle \vec{k}m|\hat{\vec{p}}|\vec{k}n\rangle \nonumber \\
 = & i\sum_{MN} C^{\vec{k}m *}_M C^{\vec{k}n}_N \sum_{\vec{R}} \langle0 M|\hat{H}|\vec{R}N\rangle \vec{R} e^{i\vec{k}\cdot\vec{R}} \nonumber \\
 + & i\epsilon_{\vec{k}m}\sum_{MN} C^{\vec{k}m *}_M C^{\vec{k}n}_N 
   \Big(\sum_{\vec{R}} \langle 0 N|\hat{\vec{r}}|\vec{R}M\rangle e^{i\vec{k}\cdot\vec{R}}\Big)^* \nonumber \\
 - & i\epsilon_{\vec{k}n}\sum_{MN} C^{\vec{k}m *}_M C^{\vec{k}n}_N \sum_{\vec{R}} 
   \langle 0 M|\hat{\vec{r}}|\vec{R}N\rangle e^{i\vec{k}\cdot\vec{R}}.
\label{eq:eq13}
\end{align}

Alternatively, the second term of the right-hand side of Eq.~\ref{eq:eq13} can be expressed as
\begin{align}
  & i\epsilon_{\vec{k}m}\sum_{MN} C^{\vec{k}m *}_M C^{\vec{k}n}_N 
    \sum_{\vec{R}} \langle\vec{R}M|\hat{\vec{r}}|0N\rangle e^{-i\vec{k}\cdot\vec{R}} \nonumber \\
= & i\epsilon_{\vec{k}m}\sum_{MN} C^{\vec{k}m *}_M C^{\vec{k}n}_N 
    \sum_{\vec{R}} \langle 0 M|(\hat{\vec{r}}-\vec{R})|\vec{R}N\rangle e^{i\vec{k}\cdot\vec{R}}, 
\end{align}
and then Eq.~\ref{eq:eq13} can be written as
\begin{align}
 & \langle \vec{k}m|\hat{\vec{p}}|\vec{k}n\rangle \nonumber \\
 = & \sum_{MN} C^{\vec{k}m *}_M C^{\vec{k}n}_N \sum_{\vec{R}} \langle0 M|\hat{H}|\vec{R}N\rangle i\vec{R} e^{i\vec{k}\cdot\vec{R}} \nonumber \\
 + & i(\epsilon_{\vec{k}m}-\epsilon_{\vec{k}n})\sum_{MN} C^{\vec{k}m *}_M C^{\vec{k}n}_N 
    \sum_{\vec{R}} \langle 0 M|\hat{\vec{r}}|\vec{R}N\rangle e^{i\vec{k}\cdot\vec{R}} \nonumber \\
 - & \epsilon_{\vec{k}m}\sum_{MN} C^{\vec{k}m *}_M C^{\vec{k}n}_N 
    \sum_{\vec{R}} \langle 0 M|\vec{R}N\rangle i\vec{R} e^{i\vec{k}\cdot\vec{R}}.
\label{eq:eq15}
\end{align}
The above equation can also be simply expressed as  
\begin{align}
 & \langle \vec{k}m|\hat{\vec{p}}|\vec{k}n\rangle \nonumber \\
 = & \sum_{MN} C^{\vec{k}m *}_M C^{\vec{k}n}_N \Big( \frac{\partial H_{MN}(\vec{k})}{\partial \vec{k}}-\epsilon_{\vec{k}m} 
   \frac{\partial S_{MN}(\vec{k})}{\partial \vec{k}} \Big) \nonumber \\
 + & i(\epsilon_{\vec{k}m}-\epsilon_{\vec{k}n})\sum_{MN} C^{\vec{k}m *}_M C^{\vec{k}n}_N 
    \sum_{\vec{R}} \langle 0 M|\hat{\vec{r}}|\vec{R}N\rangle e^{i\vec{k}\cdot\vec{R}}.
\label{eq:eq16}
\end{align}
For the special case of $m=n$, Eq.~\ref{eq:eq16} is reduced to
\begin{align}
  \langle \vec{k}n|\hat{\vec{p}}|\vec{k}n\rangle \nonumber 
 = \sum_{MN} C^{\vec{k}n *}_M C^{\vec{k}n}_N \Big( \frac{\partial H_{MN}(\vec{k})}{\partial \vec{k}} 
   - \epsilon_{\vec{k}n} \frac{\partial S_{MN}(\vec{k})}{\partial \vec{k}} \Big),  \\
\end{align}
which is exactly $\partial \epsilon_{\vec{k}n}/\partial \vec{k}$.

Finally, we demonstrate that the second term of the right-hand side of Eq.~\ref{eq:eq15} does not depend on the choice of the origin
as long as the energy eigenstates are orthogonal to each other, namely $\langle\vec{k}m|\vec{k}n\rangle = 0$ for $m\neq n$, and it is
clear that the first and third terms are origin-independent due to the relative vector $\vec{R}$. In the calculation of 
$\langle 0 M|\hat{\vec{r}}|\vec{R}N\rangle$ in two coordinate systems whose origins differ by a constant vector $\vec{d}$,
a difference can appear:
\begin{align}
\langle 0 M|\hat{\vec{r}}_2|\vec{R}N\rangle = \langle 0 M|\hat{\vec{r}}_1|\vec{R}N\rangle
+\vec{d} \langle 0 M|\vec{R}N\rangle.
\end{align}
As a result, an apparent difference for calculating $\langle \vec{k}m|\hat{\vec{p}}|\vec{k}n\rangle$ in Eq.~\ref{eq:eq15} can be found as 
\begin{align}
 i \vec{d} (\epsilon_{\vec{k}m}-\epsilon_{\vec{k}n}) \sum_{MN} C^{\vec{k}m *}_M C^{\vec{k}n}_N 
  \sum_{\vec{R}} \langle 0 M|\vec{R}N\rangle e^{i\vec{k}\cdot\vec{R}}. 
\label{eq:eq20}
\end{align}
However, following Eq.~\ref{eq:eq1}, Eq.~\ref{eq:eq20} is just the representation of $\langle\vec{k}m|\vec{k}n\rangle$ in real space 
regardless of a constant term and must be zero. 

\bibliography{refs}

\end{document}